# Channel Service Based High Speed Railway Base Station Arrangement

Chuang Zhang[*], Pingyi Fan[*], Yunquan Dong[*], Ke Xiong[*†]

*\* Department of Electronic Engineering, Tsinghua University, Beijing, P.R. China*
*† School of Computer and Information Technology, Beijing Jiaotong University, Beijing, P.R. China*
*E-mail: {zhangchuang11, dongyq08}@mails.tsinghua.edu.cn, {fpy, kxiong}@tsinghua.edu.cn*

**Abstract:** With the rapid development of high-speed railways, demands on high mobility wireless communication increase greatly. To provide stable and high data rate wireless access for users in the train, it is necessary to properly deploy base stations along the railway. In this paper, we consider this issue from the perspective of channel service which is defined as the integral of the time-varying instantaneous channel capacity. It will show that the total service quantity of each base station is a constant. In order to keep high service efficiency of the railway communication system with multiple base stations along the railway, we need to use the time division to schedule the multiple stations and allow one base station to work when the train is running close to it. In this way, we find a fact that if the ratio of the service quantity provided by each station to its total service quantity is given, the base station interval(i.e. the distance between two adjacent base stations) is a constant, regardless of the speed of the train. On the other hand, interval between two neighboring base stations will increase with the speed of the train. Furthermore, using the concept of channel service, we also analyze the transmission strategy of base stations.

## 1. Introduction

The rapid growth of high-speed railways around the world brings huge demands for wireless communications on high-speed trains, however, the current system GSM for railway(GSM-R) can only support data rate less than 200kbps, and is mainly used for signaling rather than data transmission [1]. Therefore, new technologies and systems should be developed to meet the needs of high-speed railway communication. Either simply making amendments to GSM-R or evolving the current system to a more advanced one relies on the solution of several fundamental issues. Firstly, radio signals experience dramatic loss when penetrating into or from the carriage. To deal with this problem, a two-hop architecture network has been proposed in [2] [3]. In this system, user information is firstly transmitted to the access point(AP) in the train, then the AP sends the aggregated information to the base station via an antenna on the top of the train. In this way the direct transmission from user terminals to the base station is avoided. Secondly, severe Doppler shift exists in the communication process because of high mobility, resulting in the difficulties in channel estimation. In [4], the performance of a promising scheme for high-speed railway communication--IEEE 802.11b with Doppler shift in railroad environment was analyzed. Thirdly, frequent handovers can lead to the increase of drop-offs, which significantly degrades user experiences. To reduce the handover frequency, a Radio-over-Fiber distributed antenna system was proposed in [5]. In this system, antennas are allocated along the railway and connected to a control center through fiber to increase the coverage region of a base station, thereby leading to the reduction of handover frequency. Finally, it is important to deploy the base stations properly along the railway in order to guarantee a stable service. Most of previous works considered this issue from the perspective of reducing handovers and didn't present an explicit relation between the base station interval and the speed of the train. Different from these works, we consider the base stations deployment issue from the point of view of channel service. Using this concept, we present the relation between the base station interval and the speed of the train explicitly.

The speed of the train is a key factor in the design of base station intervals, it is quite necessary to analyze their relation in details. In this work, we develop an explicit form of these two quantities using the concept of channel service. In particular, we find that when the time division scheduling is employed among the multiple base stations communication system and the ratio of the service quantity provided by each station to its total service quantity is fixed, the interval between two neighboring base stations is a constant, regardless of the speed of the train. However, if each base station is required to provide a given channel service, the interval will increase with the speed of the train.

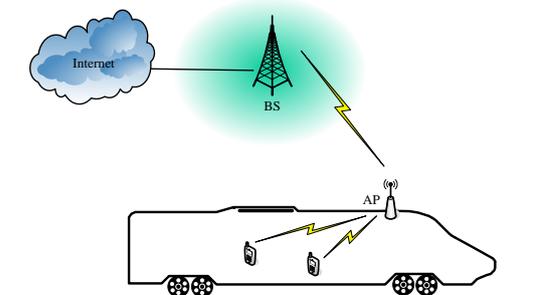

Fig.1. two-hop architecture

The rest of this paper is organized as follows: Definitions and system model are introduced in Section 2. Then, detailed discussions about base station arrangement concerning two transmission scenarios are given in Section 3. In Section 4, we use the results to analyze the transmission strategy of a base station. Finally, conclusions are given in Section 5.

## 2. System Model

### A. System Architecture

We consider the two-hop architecture model as in Fig.1. To access the network, users in the train firstly connect to the access point(AP) in the carriage, then the AP transmits the user data to a base station via the antenna on the top of the train. This structure has many advantages. Firstly, the drop-off rate could be reduced significantly. This is because the base station only needs to deal with the handoff of one AP rather than dozens of user terminals. Secondly, using the antenna on the top of the train, signals don't need to penetrate into or from the carriage, avoiding heavy energy loss. Finally, the well developed access technologies such as Wifi, WLAN can be used in the carriage to provide a stable and high speed wireless link.

It can be seen that the AP-to-base-station communication capability is the bottleneck of this two-hop architecture, which plays a key role in the service of base stations. Therefore, we mainly focus on the train-to-ground part, where the time division scheduling among multiple base stations is employed.

It is clear that there are few reflections and objections on the ways of signal propagation in high speed railway communication in most scenarios. Thus in this case it can be treated as a line of sight communication. Assume that the base station and the AP communicate through an additive white gaussian noise(AWGN) channel. At each discrete time $t$, the channel output is

$$Y_t = gX_t + Z_t, \quad (1)$$

where $Z_t \sim N(0, N_0/2)$ and independent of channel input $X_t$, $g$ is the channel gain. Given a power constraint $P_s$, the capacity of the discrete-time AWGN channel [6] is:

$$C(P_s) = \log(1 + \frac{g^2 P_s}{N_0/2}). \quad (2)$$

$\frac{g^2 P_s}{N_0/2}$ is the received signal to noise ratio(SNR).

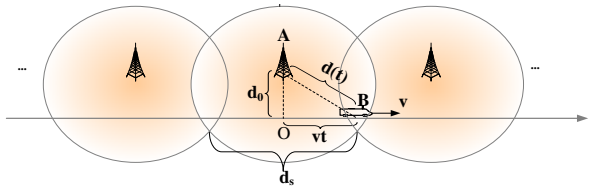

Fig.2. base station coverage model

The base station coverage model is given in Fig.2. In this figure, the base station is represented as $A$, the train is denoted as $B$. Assume that the train is moving at a constant speed $v$, the base station(which is $d_0$ distance away from the railway) sends messages to the train using constant power $P_s$. Then the received signal power on the train is $\frac{P_s}{(d(t))^\alpha}$, where $\alpha$ is path loss exponent, $d(t)$ is the distance from the base station to the train at time $t$. We establish an axis along the railway, as shown in Fig.2: Let the cross point between the railway and its vertical line through the base station be the original point, the moving direction of the train be the positive direction. Assume that the instant when the train passes the origin is time $0$. Then at a given time $t(t \in [-\infty, \infty])$, the distance $d(t)$ is $\sqrt{d_0^2 + (vt)^2}$. The received $SNR$ at time $t$ is

$$SNR = \frac{\frac{P_s}{(d_0^2+(vt)^2)^{\frac{\alpha}{2}}}}{N_0/2} \\ = \frac{2P_s}{(d_0^2+(vt)^2)^{\frac{\alpha}{2}} N_0}. \quad (3)$$

Substituting it in equation (2), we get the channel capacity at time $t$ as:

$$C(t) = \log(1 + \frac{2P_s}{(d_0^2+(vt)^2)^{\frac{\alpha}{2}} N_0}). \quad (4)$$

For convenience, we denote $\frac{2P_s}{N_0}$ as $\rho$, then equation (4) becomes

$$C(t) = \log(1 + \frac{\rho}{(d_0^2+(vt)^2)^{\frac{\alpha}{2}}}). \quad (5)$$

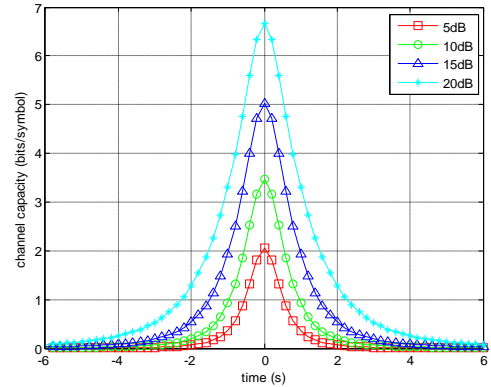

Fig.3. channel capacity versus time

The relation between channel capacity and time at different $SNR$ is illustrated in Fig.3. Note that the $SNR$ is the received signal to noise ratio chosen at time $0$ in this figure.

### B. Definition of Channel Service and Dominant Ratio

The authors in [7] proposed a new concept of channel service, which is defined as the integral of the instantaneous capacity. In fact, the channel service is the maximum amount of service that the physical layer can provide for network layer. Particularly, it is defined as

$$S(t) = \int_{-\infty}^{t} C(\tau) d\tau, \quad (6)$$

In this work, we use this concept to denote the maximum amount of service a base station can provide for the train till time $t$.

Substituting equation (5) into equation (6), we get

$$S(t) = \int_{-\infty}^{t} \log(1 + \frac{\rho}{(d_0^2 + (v\tau)^2)^{\frac{\alpha}{2}}})d\tau \qquad (7)$$

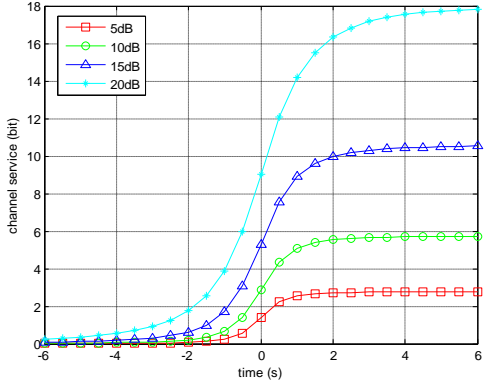

Fig.4. channel service versus time

The relationship between channel service and time is shown in Fig.4. It is seen that the curve is increasing and finally tends to a constant value for large $t$. That is to say, the total service provided by each station is a constant, which is determined by the received SNR at time 0 and the distance between base station and the railway. Its physical meaning can be explained by using the bell curve of channel capacity versus time shown in Fig.3. The channel service at time $t$ is the area of the bell before $t$, for large $t$, the channel capacity tends to $0$, and the area of the bell tends to a constant.

Note that the $C(t)$ versus $t$ curve is symmetric, we define a *dominant ratio* $\eta$ corresponding to time $t_s$ as

$$\eta = \frac{\int_{-t_s}^{t_s} C(\tau)d\tau}{\int_{-\infty}^{\infty} C(\tau)d\tau} \qquad (8)$$

$\int_{-\infty}^{\infty} C(\tau)d\tau$ is the total service of a base station, and represents the maximum amount of service the station can provide. $\int_{-t_s}^{t_s} C(\tau)d\tau$ is the service from time $-t_s$ to $t_s$, the maximum amount of service the station can provide during this time period. So, for a base station, $\eta$ is the ratio of the amount of service it can provide from time $-t_s$ to $t_s$ to its total service, thus $\eta \in [0,1]$. It can be seen that $\eta$ can be used to determine the base station interval which will be explained after the following definitions.

Define the travelling distance of the train from time $-t_s$ to $t_s$ as the *service distance* of a base station, as shown in Fig.2, which is denoted as $d_s$,

$$d_s = 2vt_s. \qquad (9)$$

Since trains move along a near linear railway, we can lay out the base stations along the track in one-dimension as shown in Fig.2. If the service distances of two adjacent base stations are $d_{s1}$ and $d_{s2}$ respectively, the interval $d_{int}$ of these two base stations is

$$d_{int} = \frac{d_{s1} + d_{s2}}{2}. \qquad (10)$$

In the following discussions, we assume that the service distances of all base stations are the same, which is $d_s$. In this case, the base station interval equals the service distance, so we will not distinguish these two quantities later, and use $d_s$ to denote both of them.

To conclude this part, we show how to use $\eta$ to determine the base station interval. If each base station is required to provide 90%, for instance, of its total service, then the service distance corresponding to this ratio is the required base station interval. Our main results regarding the relation between base station intervals and the speed of the train will be given in the next section.

## 3. Discussion on Base Station Intervals

To determine the base station interval, it is necessary to take the speed of the train into account. Our main goal in this work is to formulate the effects of the speed of the train on the decision of base station intervals. Moreover, we will analyze the relationship between channel service and the interval.

*1) Choice of Base Station Interval given a Dominant Ratio*

As is shown, *dominant ratio* can be used to determine the base station interval. The relation between base station interval and the speed of the train given a *dominant ratio* is described in the following

**Proposition 1:** *Given a dominant ratio $\eta$, the base station interval $d_s$ is a constant, regardless of the speed of the train.*

*Proof:* According to the definition of *dominant ratio*, given a constant ratio, there is:

$$\eta = \frac{\int_{-t_s}^{t_s} \log(1 + \frac{\rho}{(d_0^2 + (v\tau)^2)^{\frac{\alpha}{2}}})d\tau}{\int_{-\infty}^{\infty} \log(1 + \frac{\rho}{(d_0^2 + (v\tau)^2)^{\frac{\alpha}{2}}})d\tau}$$

$$= \frac{\int_{0}^{t_s} \log(1 + \frac{\rho}{(d_0^2 + (v\tau)^2)^{\frac{\alpha}{2}}})d\tau}{\int_{0}^{\infty} \log(1 + \frac{\rho}{(d_0^2 + (v\tau)^2)^{\frac{\alpha}{2}}})d\tau}. \qquad (11)$$

Using the substitution $x = v\tau$ for the integral, and note that $d_s = 2vt_s$, we get:

$$\eta = \frac{\int_{0}^{\frac{d_s}{2}} \log(1 + \frac{\rho}{(d_0^2 + x^2)^{\frac{\alpha}{2}}})dx}{\int_{0}^{\infty} \log(1 + \frac{\rho}{(d_0^2 + x^2)^{\frac{\alpha}{2}}})dx}. \qquad (12)$$

Change the form of equation (12), then:

$$\int_0^{\frac{d_s}{2}} \log(1+\frac{\rho}{(d_0^2+x^2)^{\frac{\alpha}{2}}})dx = \eta \int_0^{\infty} \log(1+\frac{\rho}{(d_0^2+x^2)^{\frac{\alpha}{2}}})dx. \quad (13)$$

The right side of this equation is a constant since the integral is definite, and the left side is a monotonously increasing function of $d_s$. To make the equation hold, $d_s$ must be a constant. Besides, $v$ is erased in the substitution, so it is irrelevant to $d_s$. Therefore, $d_s$ is a constant regardless of $v$. ∎

According to this proposition, if we use *dominant ratio* to determine the base station interval, then the interval does not depend on the speed of the train. This also indicates the following conclusion:

**Corollary:** *A train travelling from point $-l$ to point $l$ receives the same dominant ratio of the total service that a base station can offer no matter how fast it travels at a constant speed.*

*Proof:* The quantity service that a train can receive from point $-l$ to point $l$ is $\int_{-l/v}^{l/v} \log(1+\frac{\rho}{(d_0^2+(v\tau)^2)^{\frac{\alpha}{2}}})d\tau$, the corresponding *dominant ratio* $\eta$ is

$$\eta = \frac{\int_{-l/v}^{l/v} \log(1+\frac{\rho}{(d_0^2+(v\tau)^2)^{\frac{\alpha}{2}}})d\tau}{\int_{-\infty}^{\infty} \log(1+\frac{\rho}{(d_0^2+(v\tau)^2)^{\frac{\alpha}{2}}})d\tau}. \quad (14)$$

Use the substitution $x = v\tau$, we get

$$\eta = \frac{\int_0^{l} \log(1+\frac{\rho}{(d_0^2+x^2)^{\frac{\alpha}{2}}})dx}{\int_0^{\infty} \log(1+\frac{\rho}{(d_0^2+x^2)^{\frac{\alpha}{2}}})dx}. \quad (15)$$

Since the denominator is a constant, the nominator depends only on $l$, which is a given constant, so $\eta$ is a constant, this concludes our results. ∎

Equation (12) can be further simplified for $\alpha = 2$, which is given as:

$$\eta = \frac{\frac{d_s}{2}\ln(1+\frac{\rho}{B}) + 2A\arctan\frac{d_s/2}{A} - 2d_0\arctan\frac{d_s/2}{d_0}}{\pi(A-d_0)}, \quad (16)$$

where $A = \sqrt{\rho + d_0^2}$, $B = d_0^2 + (\frac{d_s}{2})^2$.

For $\alpha > 2$, the indefinite integral doesn't have an explicit expression. Nevertheless, this does not affect our analysis.

From equation (12), one can see that dominant ratio $\eta$ increases with the base station interval $d_s$, and the $\eta$ versus $d_s$ curve is actually concave, as shown in Fig.5. The concavity of the curve hints that given a base station interval, in order to get the maximum $\eta$ for the adjacent two base stations, the optimal policy is for each of them to provide service for only half the interval. This is easily seen from the fact that channel capacity decreases with distance in both directions along the railway.

Fig.5 also indicates that the higher the *SNR* at the original point, the more disperse the power toward the original point, the more disperse the power toward the two ends and correspondingly, the longer the trail of power distribution along the railway. Thus, for the same service distance $d_s$, smaller *SNR* actually gets higher dominant ratio $\eta$.

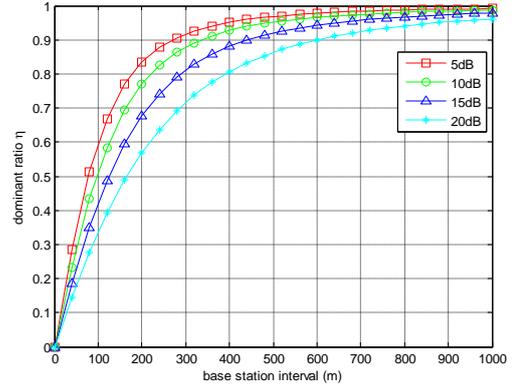

Fig.5. dominant ratio $\eta$ versus base station interval

*2) Choice of Base Station Interval given a Channel Service*

It has been shown in previous part that given a *dominant ratio*, the base station interval is irrelevant to the speed of the train. In this section, we show that if the service quantity provided by each station rather than dominant ratio is used to determine the service distance, then the base station interval will have a close relation with the speed of the train.

We first show how to use service quantity to determine the service distance: If each base station is required to provide a given amount of service in its service region, the least distance that the train travels during the transmission process is the service distance of the base station. This can be described as follows:

$$\mathbb{S} = \int_{-\frac{d_s}{2v}}^{\frac{d_s}{2v}} C(\tau)d\tau \quad (17)$$

where, $d_s$ is the service distance decided by service $\mathbb{S}$.

**Proposition 2:** *The base station interval decided by a given channel service increases with the speed of the train.*

*Proof:* Assume that the base station interval corresponding to a given channel service $\mathbb{S}$ is $d_s$, then

$$\mathbb{S} = \int_{-\frac{d_s}{2v}}^{\frac{d_s}{2v}} \log(1+\frac{\rho}{(d_0^2+(v\tau)^2)^{\frac{\alpha}{2}}})d\tau. \quad (18)$$

Use the substitution $x = v\tau$, the equation becomes

$$\mathbb{S} = \frac{2}{v}\int_0^{\frac{d_s}{2}} \log(1+\frac{\rho}{(d_0^2+x^2)^{\frac{\alpha}{2}}})dx. \quad (19)$$

Change the form of this equation, we get

$$v = \frac{2}{\mathbb{S}}\int_0^{\frac{d_s}{2}} \log(1+\frac{\rho}{(d_0^2+x^2)^{\frac{\alpha}{2}}})dx. \quad (20)$$

This equation is a function between $v$ and $d_s$, it can be easily seen from this equation that the base station interval $d_s$ increases with the speed of the train $v$. ∎

Likewise for $\alpha = 2$, there is a closed form for the integral on the right side, which is given by

$$v = \frac{2}{\mathbb{S}\ln 2}[\frac{d_s}{2}\ln(1+\frac{\rho}{B}) + 2A\arctan\frac{\frac{d_s}{2}}{A} - 2d_0\arctan\frac{\frac{d_s}{2}}{d_0}]. \quad (21)$$

where $A = \sqrt{\rho + d_0^2}$, $B = d_0^2 + (\frac{d_s}{2})^2$. For $\alpha > 2$, there is no explicit form for the integral, so we maintain the form of equation (20) for our discussion.

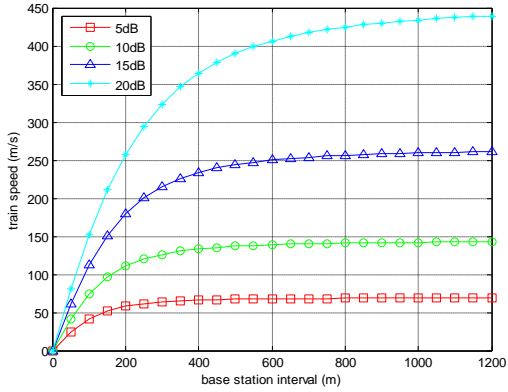

Fig.6. train speed versus base station interval

The relation between train speed $v$ and base station interval $d_s$ is shown in Fig.6. It can be seen that given a constant service, the higher the speed of the train, the longer the service distance. And for a given service distance $d_s$, in order to provide the same amount of service, it requires that the train keep a higher speed when the SNR at the original increases.

## 4. Discussion on the Transmission Strategy of Base Stations

In Section 3, we focused on the decision of base station intervals(i.e. how the train speed affects the interval). And from the above discussion, we see that the channel service is influenced by the train speed, since the channel service upper bounds the service provided by a base station, we can then use this concept to discuss the transmission strategy of a base station. We define the transmission strategy problem of base stations as follows:

A base station buffers a certain amount of data, which is equal to the total service it can provide, if it is required to provide a given ratio or amount of the buffer data to a passing train, when(i.e. at what points the train passes) should the base station start and end its transmission?

Using the results in section 3, one can get the following answers: Firstly, to transmit a certain ratio $\eta$ of the data, the distance the train travels during the transmission process has nothing to do with the speed of the train, which can be obtained by solving $d_s$ from equation (12). This is true since both the amount of data the train receives and the maximum amount of data the base station buffers decrease with $v$, as in equation (19). However, to transmit a certain amount of data, the speed of the train does affect the transmission's beginning and ending time, their relation is shown in equation (20). Intuitively, the higher the speed, the longer the distance the station needs to serve, and the earlier (later) it needs to start (end) to transmit.

## 5. Conclusion

In this work, we discussed the issue of base station deployment using the concept of channel service. Specifically, we analyzed the relation between the base station interval and the speed of the train in two scenarios. In first scenario, we guaranteed that a certain ratio of the station's maximum channel service be provided, and proved that the interval of adjacent base stations is irrelevant to the speed of the train. In the second scenario, if a given amount of service is demanded, the interval correlates with the train speed in a explicit form given by equation (20). Moreover, these results were applied in the analysis of the transmission strategy of a base station.

**Acknowledgements**

This work was partly supported by the China Major State Basic Research Development Program (973 Program) No.2012CB316100(2), National Natural Science Foundation of China(NSFC) No.61171064, the China National Science and Technology Major Project No.2010ZX03003-003 and NSFC No. 61021001.